\documentclass[%
reprint,
superscriptaddress,
amsmath,amssymb,
 aps,
prb,
floatfix,
]{revtex4-2}

\usepackage{mathbbol}

\usepackage{graphicx}
\usepackage{dcolumn}
\usepackage{bm}
\usepackage{color}
\usepackage{slashed}

\newtheorem{theorem}{Theorem}

\newtheorem{remark}{Remark}
\newtheorem{proof}{Proof}

\def\ldef{\mathrel{\mathop:}=}

\renewcommand\vec[1]{\mathbf{#1}}
\renewcommand{\Re}{\operatorname{Re}}
\renewcommand{\Im}{\operatorname{Im}}
\def\tr{\mathrm{tr}}
\def\H{\mathrm{Hess}(E_0)}

\begin{document}

\title{Zero Curvature Condition for Quantum Criticality}

\author{Chaoming Song}
\email{c.song@miami.edu}
\affiliation{%
Department of Physics, University of Miami, Coral Gables, Florida 33146, USA}%

\begin{abstract}
Quantum criticality often lies beyond the scope of the conventional Landau paradigm, and a unifying framework has yet to emerge, due in part to the wide variety of quantum orders. We propose a geometric approach to quantum phase transitions (QPTs) that shifts focus from microscopic order to the competition between non-commuting operators. This competition is encoded in the boundary geometry of their expectation values, defining a quantum observable space (QOS). We show that QPTs occur at zero-curvature points on the QOS boundary, signaling maximal commutativity and suggesting an underlying integrable structure at criticality.
\end{abstract}

\maketitle

{\it Introduction. ---}
Quantum phase transitions (QPTs) at zero temperature exhibit phenomena with no classical counterparts~\cite{vojta2003quantum,sachdev1999quantum,sachdev2000quantum,zurek2005dynamics}. While classical transitions are well described by the Landau-Ginzburg-Wilson (LGW) paradigm, quantum criticality poses deeper challenges due to the roles of entanglement and operator non-commutativity~\cite{vidal2003entanglement}. In the LGW framework, transitions are driven by spontaneous symmetry breaking of local order parameters~\cite{landau1937theory}, a picture that fails to capture many QPTs. For example, topological phase transitions involve global invariants rather than local order parameters~\cite{wen1989vacuum,wen1990topological,wen1993transitions,qi2008topological,wen2002quantum,wen2017colloquium,wen2019choreographed}. Another key case is deconfined quantum criticality, where continuous transitions emerge between competing orders, defying LGW expectations of first-order behavior or phase coexistence~\cite{senthil2004deconfined,senthil2004quantum,shyta2022frozen}.

Several frameworks have been developed to characterize QPTs beyond the LGW paradigm. Information-geometric approaches equip the manifold of quantum states with a Riemannian structure~\cite{provost1980riemannian}, using fidelity susceptibility and the quantum geometric tensor to detect criticality via singularities in ground-state overlaps~\cite{zanardi2006ground,zanardi2007information,gu2010fidelity}. Entanglement-based methods describe critical states through scale-invariant tensor networks that efficiently capture long-range quantum correlations~\cite{vidal2007entanglement}. The onset of strong correlations near criticality also leads to a sharp increase in the sign problem~\cite{mondaini2022quantum}. Exceptional points in non-Hermitian systems provide another route, signaling criticality through spectral singularities~\cite{heiss2012physics}. Despite these advances, a general and physically transparent framework remains elusive, partly due to the diversity of microscopic mechanisms. A common structural feature, often implicit, is the non-commutativity of observables.

In this Letter, we introduce a geometric framework that places non-commutativity at the center. Rather than focusing on microscopic order, we study QPTs via the geometry of expectation values of competing non-commuting operators in the Hamiltonian. These define a quantum observable space (QOS), whose boundary encodes the interplay between operators. We show that this boundary is in one-to-one correspondence with ground states and serves as a geometric generalization of \emph{Heisenberg's uncertainty principle}. Quantum critical points emerge as \emph{zero-curvature} points on the boundary, indicating maximal commutativity and potential integrability. This yields a representation-independent and experimentally accessible criterion for criticality, offering a complementary perspective beyond the LGW paradigm.

\begin{figure}
\includegraphics[width=\linewidth]{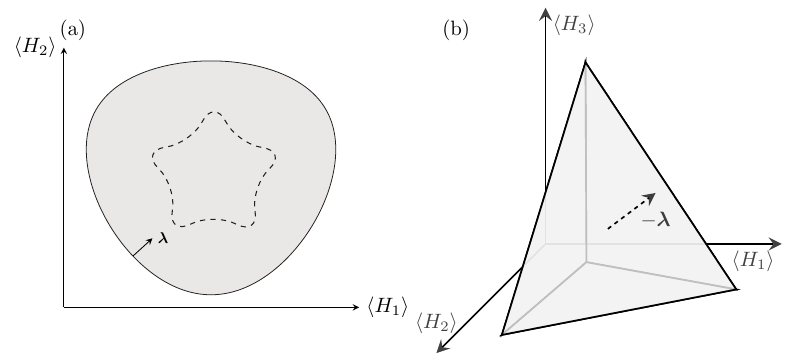}
\caption{Quantum observable space (gray region). (a) For two competing operators, shown in the \(\langle H_1 \rangle\)–\(\langle H_2 \rangle\) plane. The solid curve denotes the ground-state boundary \(\partial \mathcal{M}\), while the dashed curve represents the excited-state boundary, which is generally non-convex. (b) QOS for an integrable model with three competing operators.}
\label{fig:demo}
\end{figure}

{\it Convexity of Quantum Observable Space —} 
We consider a quantum system with Hilbert space \(\mathcal{H}\) and two semibounded Hermitian operators \(H_1, H_2 \in \mathrm{Herm}(\mathcal{H})\), following the framework of Ref.~\cite{song2023quantum}. These define a parametric Hamiltonian family
\begin{equation}\label{eq:H}
H(\lambda_1, \lambda_2) = \lambda_1 H_1 + \lambda_2 H_2,
\end{equation}
with real parameters \(\lambda_1, \lambda_2\) that govern the competition between generally non-commuting terms. The associated pure-state quantum observable space (QOS) is defined as
\begin{equation}
\mathcal{M} \ldef \left\{ \left(\langle H_1 \rangle_\psi, \langle H_2 \rangle_\psi \right) \colon |\psi\rangle \in \mathcal{H},\; \|\psi\| = 1 \right\},
\notag
\end{equation}
i.e., the set of all expectation value pairs over normalized pure states, which is also known as the joint numerical range~\cite{gustafson1997numerical,bonsall1971numerical}. For finite-dimensional \(\mathcal{H}\), \(\mathcal{M}\) is a compact semialgebraic subset of \(\mathbb{R}^2\), with a non-trivial boundary curve \(\partial \mathcal{M}\), which plays a central role in our analysis (see Fig.~\ref{fig:demo}a).

A key property of \(\mathcal{M}\) is its convexity, as first shown by Brickman~\cite{brickman1961field} as an extension of the Hausdorff–Toeplitz theorem~\cite{toeplitz1918algebraische,hausdorff1919wertvorrat}. To see this, we observe that for any pair of states \(|\psi_1\rangle\), \(|\psi_2\rangle\), the interpolating superposition
\(
|\psi(p)\rangle = \sqrt{p}\,|\psi_1\rangle + \sqrt{1 - p}\, e^{i\theta} |\psi_2\rangle
\)
generates a continuous path connecting the corresponding points in \(\mathcal{M}\), thereby tracing the entire segment between them for a suitable choice of \(\theta\) and all \(p \in [0,1]\) (see Appendix A). This implies that \(\mathcal{M}\) is convex. An alternative argument based on Ehrenfest’s theorem and monotonicity also supports this conclusion~\cite{sorensen2001entanglement}.

This construction generalizes to any number \(n\) of semibounded Hermitian operators \(\{H_1, \ldots, H_n\}\), defining the Hamiltonian family
\begin{equation} \label{eq:Hn}
H = \lambda_1 H_1 + \cdots + \lambda_n H_n.
\end{equation}
Here, \(\lambda_i\) are real if \(H_i\) is bounded, and nonnegative if \(H_i\) is only semibounded from below. The pure-state QOS generalizes to an \(n\)-dimensional set
\begin{equation}
\mathcal{M} \ldef \left\{ \left(\langle H_1 \rangle_\psi, \ldots, \langle H_n \rangle_\psi \right) \colon |\psi\rangle \in \mathcal{H},\; \|\psi\| = 1 \right\}.
\notag
\end{equation}
For \(n > 2\), however, the set \(\mathcal{M}\) is no longer guaranteed to be convex, with counterexamples such as the Bloch sphere. Nonetheless, convexity is restored when one extends to the mixed-state QOS,
\[
\overline{\mathcal{M}} \ldef \left\{ \left( \mathrm{Tr}(\rho H_1), \ldots, \mathrm{Tr}(\rho H_n) \right) \colon \rho \in \mathcal{D},\; \mathrm{Tr}(\rho) = 1 \right\},
\]
where \(\mathcal{D} = \{ \rho \in \mathrm{Herm}(\mathcal{H}) \colon \rho \ge 0 \}\) is the space of density matrices. Since \(\mathcal{D}\) is convex and the trace is linear, \(\overline{\mathcal{M}}\) is convex by construction, as has been extensively studied in the theory of operator algebras~\cite{alfsen2012state}. Clearly, \(\mathcal{M} \subseteq \overline{\mathcal{M}}\). Remarkably, these two sets actually coincide under mild conditions. If the observables \(\{H_i\}\) are generic, i.e., their common eigenspaces form a set of measure zero, then 
$\mathcal{M} = \overline{\mathcal{M}}$ (see Appendix B).
This requires the dimension of \(\mathcal{H}\) to be no smaller than \(\lceil n/2 \rceil + 1\), which always holds for \(n = 2\). In physically relevant settings, such as many-body systems in the thermodynamic limit or field-theoretic models with infinite-dimensional \(\mathcal{H}\), the Hilbert space is sufficiently large that such a generic choice is almost surely satisfied. In such cases, the pure-state QOS is convex and coincides with its mixed-state counterpart. Throughout this work, we therefore identify \(\mathcal{M} = \overline{\mathcal{M}}\), and thus their convexity, unless stated otherwise.

{\it Geometric Uncertainty Principle ---}
To analyze the structure of the quantum observable space \(\mathcal{M}\), we consider the energy functional
\begin{equation}\label{eq:E}
    E[\psi] = \lambda_1 \langle H_1 \rangle + \cdots + \lambda_n \langle H_n \rangle,
\end{equation}
with \(\langle H_i \rangle \equiv \langle \psi | H_i | \psi \rangle\), and \(|\psi\rangle \in \mathcal{H}\), \(\|\psi\| = 1\). The variational condition \(\delta E[\psi] = 0\) implies that stationary points correspond to eigenstates of the Hamiltonian~\eqref{eq:Hn}, satisfying
\begin{equation} \label{eq:var}
    \lambda_1\, \delta \langle H_1 \rangle + \cdots + \lambda_n\, \delta \langle H_n \rangle = 0.
\end{equation}

The expectation values of eigenstates \(|\psi_e(\boldsymbol{\lambda})\rangle\), for fixed \(\boldsymbol{\lambda} = (\lambda_1, \ldots, \lambda_n)\), form a singular set
\begin{equation}
    \mathcal{M}_e = \left\{ (\langle H_1 \rangle_e, \ldots, \langle H_n \rangle_e) \colon |\psi_e\rangle,\; \|\psi_e\| = 1 \right\} \subseteq \mathcal{M},
\notag
\end{equation}
where the Jacobian of the map \(|\psi\rangle \mapsto (\langle H_1 \rangle, \ldots, \langle H_n \rangle)\) becomes rank-deficient~\cite{hartshorne2013algebraic,song2023quantum}. Hence, the variational condition defines a critical variety, with \(\boldsymbol{\lambda}\) as the local normal vector to \(\mathcal{M}_e\).

For finite-dimensional \(\mathcal{H}\), \(\mathcal{M}_e\) forms an \((n-1)\)-dimensional algebraic variety defined by a bivariate polynomial~\cite{song2023quantum}. Of particular interest is the subset of ground states,
\begin{equation}
    \mathcal{M}_0 = \left\{ (\langle H_1 \rangle_0, \ldots, \langle H_n \rangle_0) \colon |\psi_0\rangle,\; \|\psi_0\| = 1 \right\} \subseteq \mathcal{M}_e,
\notag
\end{equation}
where \(|\psi_0\rangle\) is the ground state. For semibounded operators, this corresponds to the lowest eigenstate. For bounded operators, either extremum may define the ground state depending on the sign of the Hamiltonian.

We find that the boundary of QOS coincides with this ground-state space (see Appendix C),
\begin{equation}
    \partial \mathcal{M} = \mathcal{M}_0,
\label{eq:bound}
\end{equation}
as a consequence of the supporting hyperplane theorem for convex sets~\cite{boyd2004convex}, together with the variational inequality
\begin{equation}
\lambda_1 \langle H_1 \rangle + \cdots + \lambda_n \langle H_n \rangle \geq \lambda_1 \langle H_1 \rangle_0 + \cdots + \lambda_n \langle H_n \rangle_0,
\notag
\end{equation}
Equation~\eqref{eq:bound} defines a geometric generalization of Heisenberg's uncertainty principle. For example, the standard relation \(\langle p^2 \rangle \langle x^2 \rangle \geq \hbar^2 / 4\) defines a QOS for the harmonic oscillator, with the saturation curve lying precisely on the boundary and realized by the ground state. Thus, the geometry of \(\partial \mathcal{M}\) encodes fundamental quantum constraints on jointly observable quantities.

\begin{figure}
 \includegraphics[width=1\linewidth]{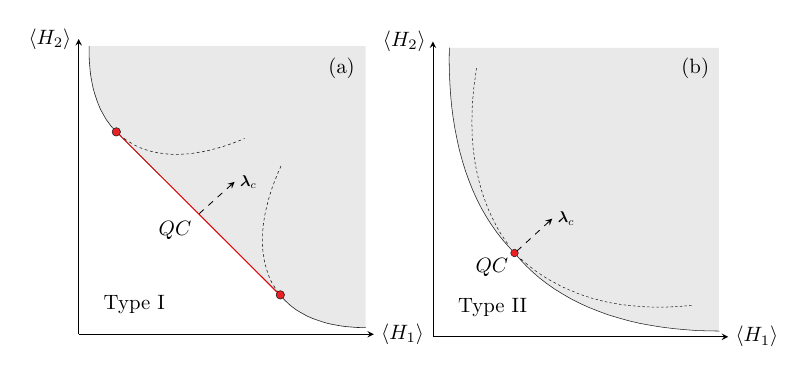}
 \caption{ (a) Type I and (b) Type II quantum critical points (QC). The dashed curves represent the first excited states. 
 }. 
 \label{fig:curvature}
\end{figure}

{\it Zero Curvature Condition.---} 
The convexity of \(\mathcal{M}\) implies that the Gaussian curvature of its boundary \(\partial \mathcal{M}\) is non-negative:
\begin{equation}\label{eq:kappa}
    \kappa \geq 0,
\end{equation}
and more generally, the second fundamental form is positive semidefinite for \(n > 2\). This holds only for the ground-state manifold, reflecting its fundamental role in QPTs as distinct from excited-state behavior relevant to finite-temperature transitions.

In the thermodynamic limit, after rescaling observables by system size, \(\partial \mathcal{M}\) may develop non-analyticities signaling QPTs. However, convexity restricts the types of singularities: for example, cusps may appear in excited-state geometry but not for ground states. The QPT appears as a singularity on \(\partial \mathcal{M}\), where the curvature vanishes,
\begin{equation}\label{eq:cur}
    \kappa_c = 0.
\end{equation}

Two types of zero-curvature singularities may arise. Type I, illustrated in Fig.~\ref{fig:curvature}a, features a gap closing between two points \((\langle H_1 \rangle^*, \langle H_2 \rangle^*)\) and \((\langle H_1 \rangle^{**}, \langle H_2 \rangle^{**})\), connected by a straight segment on \(\partial \mathcal{M}\). This corresponds to a first-order transition with discontinuities \(\Delta \langle H_i \rangle\) in order parameters and typically \(\Delta E \neq 0\), though \(\Delta E = 0\) may occur under degeneracy. The common normal direction \(\boldsymbol\lambda_c = (\lambda_{1c}, \lambda_{2c})\) defines a single critical point. In contrast, Type II transitions, shown in Fig.~\ref{fig:curvature}b, occur at an isolated point with gap closing and continuous order parameters, but diverging derivatives, analogous to second-order transitions. In both cases, the zero-curvature condition \(\kappa = 0\) marks the criticality.

This geometry connects to the Ehrenfest classification. Setting \(\lambda_1 > 0\), the coupling \(g = \lambda_2/\lambda_1\) yields from Eq.~\eqref{eq:E} and the variational condition~\eqref{eq:var}:
\(
\langle H_1 \rangle = E(g) - g E'(g), \quad \langle H_2 \rangle = E'(g),
\)
via the Hellmann–Feynman theorem. The curvature becomes
\begin{align}
\kappa = -\frac{1}{(g+1)^{3/2} E''(g)},
\notag
\end{align}
showing that \(\kappa = 0\) implies a divergent second derivative. For non-degenerate ground states, perturbation theory gives
\(
E''(g) = 2 \sum_{k > 0} \frac{|\langle k | H_2 | 0 \rangle|^2}{E_0 - E_k} \leq 0,
\)
consistent with \(\kappa \geq 0\), and vanishing as the gap \(\Delta = E_1 - E_0\) closes. In fact, \(\kappa \sim \Delta\) near criticality. Our convexity result applies even in degenerate cases, extending beyond perturbative limits.

Type I and Type II transitions broadly correspond to first- and second-order phase transitions, respectively, but differ fundamentally from their classical counterparts. While classical transitions are typically driven by symmetry breaking, quantum transitions in our framework arise from the competition between non-commuting operators, imposing geometric constraints absent in classical theories. In particular, Type I transitions occur only at fixed normal directions
\(\boldsymbol\lambda_c\). As a result, they differ from classical first-order transitions, where coexistence spans a finite parameter region and involves mixed states, as described by Landau theory. In the quantum case, the ground state remains pure at zero temperature, and coexistence reflects degeneracy between distinct ground states. The system may interpolate between these extremal states with a fixed $(\boldsymbol\lambda_c)$, but no extended coexistence region exists in the parameter space.

Furthermore, in systems involving more than two competing operators, the zero-curvature condition extends naturally to higher dimensions. In such cases, it defines critical geometries that delineate the phase boundaries of multidimensional phase diagrams.

{\it Local Integrability.---}
To better understand the zero-curvature condition, we investigate the geometry of \(\mathcal{M}\) in an integrable system. Let \(H_1, \ldots, H_n\) be a set of linearly independent, mutually commuting Hermitian operators acting on a finite-dimensional Hilbert space \(\mathcal{H}\), with \(\dim \mathcal{H}=n\). Without loss of generality, we may set \(H_n = I\) and \(\lambda_n = -E\), so that the Schrödinger equation takes the form \(H|\psi\rangle = 0\).

Since the operators commute, they can be simultaneously diagonalized as
\(
H_i = \mathrm{diag}(\Lambda_{i1}, \ldots, \Lambda_{in}),
\)
where \(\Lambda_{ij}\) is the \(j\)-th eigenvalue of \(H_i\). For any normalized state \(|\psi\rangle = \sum_j \psi_j |j\rangle\), the expectation values are
\(
\langle \psi | H_i | \psi \rangle = \sum_{j=1}^n \Lambda_{ij} |\psi_j|^2.
\)
The matrix \(\Lambda = (\Lambda_{ij})\) is invertible due to the linear independence of the operators \(\{H_i\}\). Defining its inverse \(\tilde{\Lambda} = \Lambda^{-1}\), we obtain the relation
\(
\sum_{j=1}^n \tilde{\Lambda}_{ij} \langle \psi | H_j | \psi \rangle = |\psi_i|^2 \geq 0.
\)
Substituting \(\langle \psi | H_n | \psi \rangle = \langle \psi | \psi \rangle = 1\), we derive a set of \(n\) linear inequalities:
\begin{equation} \label{eq:simplex}
\sum_{j=1}^{n-1} \tilde{\Lambda}_{ij} \langle H_j \rangle + \tilde{\Lambda}_{in} \geq 0,
\end{equation}
where \(\langle H_j \rangle = \langle \psi | H_j | \psi \rangle\). These inequalities define a simplex in \((n-1)\)-dimensional space, and the boundary \(\partial \mathcal{M}\) is given by the locus where equality holds in Eq.~\eqref{eq:simplex}.

For any \(n < \dim \mathcal{H}\), the geometry of \(\mathcal{M}\) is obtained by projecting the \(\dim \mathcal{H}\)-dimensional simplex onto an \(n\)-dimensional subspace, resulting in a convex polytope. For example, when \(H_1\) and \(H_2\) commute in Eq.~\eqref{eq:H}, the corresponding QOS is a polygon. The Gaussian curvature \(\kappa\) of this polygon vanishes almost everywhere on the polygonal boundary, except at singular points where the boundary is non-differentiable. By contrast, for non-commuting operators, one typically observes a smooth, curved boundary. Therefore, the zero-curvature condition implies that the system near the critical point approximates an integrable limit, where the competing operators \(H_1\) and \(H_2\) are nearly commuting.

\textit{Examples.---}
We now apply our theory to several representative systems. One of the simplest examples of a QPT is the one-dimensional transverse field Ising model (TFIM)~\cite{pfeuty1970one}, described by the Hamiltonian
\begin{equation}
H_{\mathrm{TFIM}} = -J \sum_i Z_i Z_{i+1} - h \sum_i X_i,
\notag
\end{equation}
where \(X_i\) and \(Z_i\) are Pauli matrices acting on site \(i\), and \(J\) and \(h\) denote the spin interaction strength and the transverse field, respectively. In our setting, we identify \(H_1 = - \frac{1}{L} \sum_i Z_i Z_{i+1}\) and \(H_2 = - \frac{1}{L} \sum_i X_i\), where \(L\) is the number of sites.

\begin{figure}[t]
 \includegraphics[width=1\linewidth]{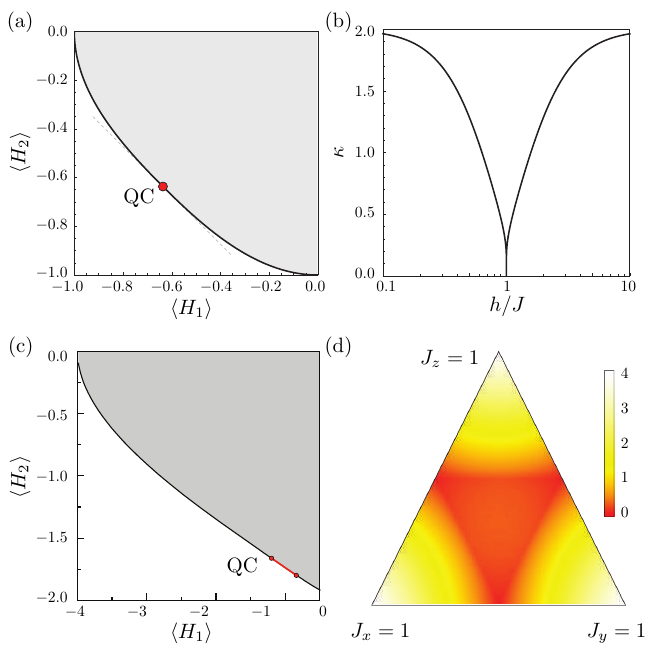}
 \caption{(a) The quantum observable space and (b) curvature for TFIM. (c) Quantum observable space for the \(2\)D three-state quantum Potts model, with data extracted from Ref.~\cite{ding2017monte}. (d) Curvature heatmap for the Kitaev honeycomb model.}
 \label{fig:example}
\end{figure}

The TFIM follows the conventional LGW paradigm of symmetry breaking, with magnetization \(m(g) \equiv -\langle H_2 \rangle\) serving as the order parameter, where \(g \equiv h/J\) is the dimensionless coupling. The model exhibits Kramers-Wannier duality, which interchanges \(H_1\) with \(H_2\) and \(J\) with \(h\), implying that \(\langle H_1 \rangle = -m(1/g)\) gives the magnetization of the dual system~\cite{pfeuty1970one}. Figure~\ref{fig:example}a plots \(\langle H_1 \rangle\) versus \(\langle H_2 \rangle\), tracing the boundary of the quantum observable space \(\mathcal{M}\). A quantum phase transition occurs at the critical point \(g_c = 1\) (marked by the red dot). The Gaussian curvature \(\kappa\) can be computed using the exact magnetization in the thermodynamic limit~\cite{pfeuty1970one}, yielding
\begin{equation}
\kappa = \frac{\pi g^2 (g+1)\left(g^2+1\right)^{-3/2}}{ \left(g^2+1\right) K\left(\frac{4g}{(g+1)^2}\right)-(g+1)^2 E\left(\frac{4g}{(g+1)^2}\right)},
\notag
\end{equation}
where \(K(x)\) and \(E(x)\) are complete elliptic integrals of the first and second kinds. Figure~\ref{fig:example}b shows the curvature as a function of \(g\), vanishing at the critical point in agreement with Eq.~\eqref{eq:cur}. Additionally, the curvature satisfies the self-duality relation \(\kappa(g) = \kappa(1/g)\).

The TFIM is a special case of the quantum \(q\)-state Potts model with \(q = 2\). In 1D for \(q > 4\) or in 2D for \(q > 2\), the quantum Potts model undergoes a first-order phase transition~\cite{solyom1981renormalization}, which corresponds to a Type-I singularity in our classification. To verify this, Fig.~\ref{fig:example}c shows \(\langle H_1 \rangle\) versus \(\langle H_2 \rangle\) for the ground state, based on numerical data from Ref.~\cite{ding2017monte} (see Appendix E). The presence of a Type-I singularity is clearly visible.

While the two examples above fall within the Landau framework, we now turn to QPTs beyond the LGW paradigm. As a non-Landau example, we consider the Kitaev honeycomb model~\cite{kitaev2006anyons}, defined on a hexagonal lattice with anisotropic bond-dependent interactions:
\begin{equation}
H_{\mathrm{KH}} = -J_x \sum_{\langle ij \rangle_x} \sigma_i^x \sigma_j^x 
                 -J_y \sum_{\langle ij \rangle_y} \sigma_i^y \sigma_j^y 
                 -J_z \sum_{\langle ij \rangle_z} \sigma_i^z \sigma_j^z,
\notag
\end{equation}
where \(\langle ij \rangle_\alpha\) denotes nearest-neighbor pairs along the \(\alpha = x, y, z\) directions. The model is exactly solvable and exhibits a topological QPT between gapped and gapless spin-liquid phases. The phase boundary is given by \(|J_\alpha| = |J_\beta| + |J_\gamma|\) for all permutations of \(\alpha \neq \beta \neq \gamma\)~\cite{feng2007topological,chen2008exact}.

Here, the Hamiltonian involves three competing terms, and the QOS becomes a three-dimensional set with expectation values \(\langle H_\alpha \rangle = - \sum_{\langle ij \rangle_\alpha} \langle \sigma_i^\alpha \sigma_j^\alpha \rangle\). The boundary \(\partial \mathcal{M}\) forms a two-dimensional surface. We compute the Gaussian curvature analytically (see Appendix D),
\begin{equation}
\kappa = 2 \left( \frac{1}{\mathrm{Vol}(\text{BZ})^2} \iint_\text{BZ} \left| \vec{u}(\vec{k}) \wedge \vec{u}(\vec{k}') \right|^2 \, d\vec{k} \, d\vec{k}' \right)^{-1},
\notag
\end{equation}
where \(\text{BZ}\) is the Brillouin zone, and \(\vec{u}(\vec{k}) = |f|^{-3/2} \Im(f^* \vec{\Phi})\), with \(f = \vec{J} \cdot \vec{\Phi}\) and \(\vec{\Phi}_i = e^{i \vec{k} \cdot \vec{a}_i}\). The curvature is non-negative, consistent with convexity. Figure~\ref{fig:example}d shows \(\kappa\) on the plane \(J_x + J_y + J_z = 1\), vanishing precisely on the analytically known phase boundary. This behavior can be understood analytically: at the phase boundary, two Dirac cones merge, leading to a quadratic band touching. The resulting dispersion causes the curvature integral to diverge, indicating vanishing curvature and thus confirming the zero-curvature condition for the topological transition.

{\it Discussion.---}
We have introduced a geometric framework for QPTs that applies uniformly to both Landau and non-Landau types. In our approach, the competition between non-commuting operators defines a QOS, whose boundary encodes ground state information. Quantum criticality manifests as a singularity of this boundary, specifically at points of vanishing Gaussian curvature. We showed that both first- and second-order QPTs are associated with such zero-curvature points, albeit with distinct geometric signatures: discontinuous transitions correspond to flat segments, while continuous transitions are marked by smooth but non-analytic points where the curvature vanishes.

The emergence of zero curvature signals a local increase in commutativity among competing terms in the Hamiltonian and may reflect an underlying enhancement of symmetry. This observation resonates with results from conformal field theory, where many critical points admit integrable deformations under relevant perturbations~\cite{zamolodchikov1989integrable}. Our findings suggest that quantum critical points may support effective integrable field theories in their vicinity, even when the underlying microscopic models are not integrable. Importantly, integrable systems near criticality often display strong entanglement and rich structure. Indeed, this geometric picture is consistent with the entropic formulation of the c-theorem~\cite{casini2007c}, which asserts that entanglement entropy is maximized at conformal fixed points and decreases monotonically along renormalization group flows away from criticality.

While our framework shares conceptual affinity with information-geometric approaches such as Provost–Vallee metric~\cite{provost1980riemannian} and the fidelity susceptibility~\cite{zanardi2006ground, gu2010fidelity}, it differs operationally: our curvature is defined directly in the space of observable expectation values, making it experimentally accessible and free from coordinate ambiguity. Likewise, tensor-network approaches~\cite{vidal2007entanglement} that capture scale-invariant entanglement at criticality may allow for a complementary geometric interpretation. In particular, the entanglement structure near criticality may be reflected in the shape and properties of the QOS boundary, offering a new lens on renormalization flow and universality.

This approach opens several promising directions for future work. One is the systematic classification of phase boundaries using algebro-geometric invariants of the QOS. In systems with more than two competing operators, the critical geometry may extend beyond isolated points or flat segments. While our analysis of the Kitaev honeycomb model demonstrates such a structure in three dimensions, a complete geometric characterization of criticality in higher-dimensional observable spaces remains an open challenge. Another intriguing direction is the potential connection to critical slowing down, where vanishing curvature may correspond to diverging timescales in the Kibble–Zurek mechanism. More broadly, our results suggest that quantum criticality can be understood as a geometric constraint arising from the non-commutativity of observables, offering a unified perspective on strongly correlated phases beyond the reach of traditional symmetry-based classifications.

\begin{acknowledgments}
We would like to thank Klaus Mølmer for the helpful correspondence.
\end{acknowledgments}

\appendix

\section{Convexity of \(\mathcal{M}\) for \(n = 2\)}

We prove that the quantum observable space (QOS)
\[
\mathcal{M}
:=\bigl\{(\langle H_1\rangle_\psi,\langle H_2\rangle_\psi)\colon|\psi\rangle\in\mathcal H,\ \|\psi\|=1\bigr\}
\subset\mathbb R^2
\]
is convex for any pair of Hermitian operators \(H_1, H_2\) acting on a Hilbert space \( \mathcal{H} \).

\begin{proof}
Let \( (x_1, y_1), (x_2, y_2) \in \mathcal{M} \) arise from normalized pure states \( |\psi_1\rangle, |\psi_2\rangle \), with
\[
x_i = \langle \psi_i | H_1 | \psi_i \rangle, \quad
y_i = \langle \psi_i | H_2 | \psi_i \rangle, \quad i = 1, 2.
\]
For any \( p \in [0,1] \), define the interpolating state
\[
|\psi(p)\rangle
= \sqrt{p}\, |\psi_1\rangle
+ \sqrt{1-p}\, e^{i\theta}\, |\psi_2\rangle.
\]
Let
\[
(x(p), y(p))
:= \left( \langle \psi(p) | H_1 | \psi(p) \rangle, \,
           \langle \psi(p) | H_2 | \psi(p) \rangle \right).
\]
Define:
\[
\begin{aligned}
&\Delta x := x_2 - x_1, \quad
\Delta y := y_2 - y_1, \quad
\Delta n := y_2 x_1 - x_2 y_1, \\
&x_{12} := \langle \psi_1 | H_1 | \psi_2 \rangle, \quad
y_{12} := \langle \psi_1 | H_2 | \psi_2 \rangle, \quad
n_{12} := \langle \psi_1 | \psi_2 \rangle.
\end{aligned}
\]
Expanding \( x(p) \) and \( y(p) \) gives:
\[
\begin{aligned}
x(p) &= p\, x_1 + (1-p)\, x_2 + 2 \sqrt{p(1-p)}\, \Re\left( e^{i\theta} x_{12} \right), \\
y(p) &= p\, y_1 + (1-p)\, y_2 + 2 \sqrt{p(1-p)}\, \Re\left( e^{i\theta} y_{12} \right).
\end{aligned}
\]

Now choose the phase angle:
\[
\theta := \arg\left( x_{12}\, \Delta y - y_{12}\, \Delta x - n_{12}\, \Delta n \right) + \tfrac{\pi}{2}.
\]
With this choice, a direct calculation shows:
\[
\left( x(p) - x_1 \right) \Delta y
= \left( y(p) - y_1 \right) \Delta x,
\]
so the point \( (x(p), y(p)) \) lies on the line segment between \( (x_1, y_1) \) and \( (x_2, y_2) \). In particular, there exists a function \( q(p) \in [0,1] \) such that
\[
(x(p), y(p)) = (1 - q(p)) (x_1, y_1) + q(p)\, (x_2, y_2).
\]
Since \( q(p) \) is continuous and satisfies \( q(1) = 0 \), \( q(0) = 1 \), the image of \( \psi(p) \) as \( p \) varies over \([0,1]\) sweeps the full line segment between the two points. Therefore, every point on the segment lies in \( \mathcal{M} \), and \( \mathcal{M} \) is convex.
\end{proof}

\begin{remark}[Convexity for \(n = 2\) vs.\ \(n > 2\)]
The set \(\mathcal{M}\) is also known as the \emph{joint numerical range} of \(H_1\) and \(H_2\)~\cite{gustafson1997numerical}, and is known to be convex for \(n = 2\). This result is a nontrivial extension of the classical Hausdorff–Toeplitz theorem, which states that the numerical range \( \{ \langle \psi | A | \psi \rangle \in \mathbb{C} \} \) of a bounded operator \( A \) is convex~\cite{toeplitz1918algebraische,hausdorff1919wertvorrat}.

The two-dimensional case corresponds to the \emph{joint numerical range} of \(H_1\) and \(H_2\), whose convexity was first rigorously established by Brickman~\cite{brickman1961field}, and is now a classical result in matrix analysis (see also~\cite{gustafson1997numerical,bonsall1971numerical}). However, for \(n \geq 3\), the joint numerical range is not convex in general. A simple counterexample is provided by the three Pauli matrices on \(\mathbb{C}^2\), whose joint pure-state expectation values trace the surface of the Bloch sphere, which is not convex. Nonetheless, as we will show in the next section, convexity can be recovered under mild conditions.
\end{remark}

\section{Equality \(\mathcal{M} = \overline{\mathcal{M}}\) and Convexity of \(\mathcal{M}\) for \(n \geq 2\)}

\begin{theorem}
\label{thm:convM_equals_mixed}
Let $\mathcal{H}$ be a finite-dimensional complex Hilbert space, and let $H_1, \dots, H_n \in \mathrm{Herm}(\mathcal{H})$ be Hermitian operators. Define:

\begin{itemize}
  \item The pure-state QOS:
  \[
    \mathcal{M} := \left\{ 
      \left( \langle \psi | H_1 | \psi \rangle, \dots, \langle \psi | H_n | \psi \rangle \right)
      \;\middle|\; |\psi\rangle \in \mathcal{H},\; \|\psi\| = 1
    \right\}.
  \]

  \item The mixed-state QOS:
  \[
    \overline{\mathcal{M}} := 
    \left\{ 
      \left( \mathrm{Tr}(\rho H_1), \dots, \mathrm{Tr}(\rho H_n) \right)
      \;\middle|\; \rho \in \mathcal{D}(\mathcal{H})
    \right\},
  \]
  where $\mathcal{D}(\mathcal{H}) := \{ \rho \in \mathrm{Herm}(\mathcal{H}) \mid \rho \ge 0,\; \mathrm{Tr}(\rho) = 1 \}$ is the set of density matrices.

\end{itemize}

Then:
\[
    \overline{\mathcal{M}} = \mathrm{conv}(\mathcal{M}).
\]
\end{theorem}

\begin{proof}
\emph{Pure states lie in $\overline{\mathcal{M}}$.}  
For any normalized vector $|\psi\rangle \in \mathcal{H}$, the rank-one projector $\rho := |\psi\rangle\langle\psi|$ is a valid density matrix. For such $\rho$, we have
\[
  \mathrm{Tr}(\rho H_i) = \langle \psi | H_i | \psi \rangle,
  \quad \text{for each } i = 1, \dots, n,
\]
so every point in $\mathcal{M}$ lies in $\overline{\mathcal{M}}$.

\emph{Mixed states are convex combinations of pure states.}  
Every density matrix $\rho \in \mathcal{D}(\mathcal{H})$ has a spectral decomposition
\[
  \rho = \sum_{j=1}^r p_j\, |\psi_j\rangle \langle \psi_j|,
  \quad \text{with } p_j \ge 0,\; \sum_j p_j = 1,\; \|\psi_j\| = 1.
\]
Then the expectation values are
\[
  \mathrm{Tr}(\rho H_i) = \sum_j p_j\, \langle \psi_j | H_i | \psi_j \rangle,
  \quad \text{for each } i = 1, \dots, n,
\]
so
\[
  \left( \mathrm{Tr}(\rho H_1), \dots, \mathrm{Tr}(\rho H_n) \right)
  = \sum_j p_j \left( \langle \psi_j | H_1 | \psi_j \rangle, \dots, \langle \psi_j | H_n | \psi_j \rangle \right),
\]
which is a convex combination of points in $\mathcal{M}$. Hence:
\[
  \overline{\mathcal{M}} \subseteq \mathrm{conv}(\mathcal{M}).
\]

\emph{Convex combinations of points in $\mathcal{M}$ come from mixed states.}  
Conversely, any convex combination of pure-state expectation values corresponds to a convex combination of rank-one projectors, and hence defines a valid density matrix. Therefore, any point in $\mathrm{conv}(\mathcal{M})$ arises from some $\rho \in \mathcal{D}(\mathcal{H})$.

\medskip

Combining both inclusions:
\[
    \overline{\mathcal{M}} = \mathrm{conv}(\mathcal{M}).
\]
\end{proof}

The convex geometry of the mixed-state QOS \( \overline{\mathcal{M}} \), defined as the image of the normal state space under a collection of observables, has been extensively studied in the theory of operator algebras~\cite{alfsen2012state}. Within this framework, the set \( \overline{\mathcal{M}} \) is convex by construction, and its exposed faces, support functionals, and orientation structure have been thoroughly characterized. However, the question of when \( \overline{\mathcal{M}} \) coincides with the pure-state image \( \mathcal{M} \), that is, when the full convex set can be reconstructed from pure-state expectations, remains largely unaddressed. We now present a geometric criterion showing that this equivalence holds generically under a mild condition:

\begin{theorem}[Pure-State Realizability]
\label{thm:M_equals_convM}
Let $\mathcal{H} = \mathbb{CP}^{d-1}$ be a $d$-dimensional complex Hilbert space, and let $H_1, \dots, H_n \in \mathrm{Herm}(\mathcal{H})$ be $n$ Hermitian operators. Define the map
\[
  \Phi_{\mathrm{pure}} \colon \mathcal{H}  \to \mathbb{R}^n, \quad
  [\psi] \mapsto \bigl( \langle \psi | H_1 | \psi \rangle, \dots, \langle \psi | H_n | \psi \rangle \bigr),
\]
and the image set of pure-state expectations
\[
  \mathcal{M} := \mathrm{Im}(\Phi_{\mathrm{pure}}).
\]
Let $\overline{\mathcal{M}} := \mathrm{conv}(\mathcal{M})$ be the mixed-state expectation set.

Assume the following: \textbf{(Eigenstate Non-Dominance)} The set
\begin{center}
  \[
    \Sigma := \left\{ [\psi] \in \mathcal{H} \;\middle|\; 
        \exists\, \vec{\lambda} \in \mathbb{R}^n \setminus \{0\}, \;
        \left( \sum_{i=1}^n \lambda_i H_i \right) |\psi\rangle = \lambda |\psi\rangle
    \right\}
  \]
\end{center}
has \emph{empty interior} in $\mathcal{H}$. Then the pure-state expectation set is convex:
\[
    \mathcal{M} = \overline{\mathcal{M}}.
\]
\end{theorem}

\noindent
In other words, if $d$ is large relative to $n$, and if $H_1,\dots,H_n$ are chosen generically, then all mixed-state points can be compressed into a single pure state.
\begin{proof}
We proceed in several steps.

\paragraph{Rank-deficient points are eigenstates of a combination.}

Let $[\psi] \in \mathbb{CP}^{d-1}$, and let $K_i := H_i - \langle \psi | H_i | \psi \rangle \mathrm{I}$.  
Then the differential of $\Phi_{\mathrm{pure}}$ at $[\psi]$ acts on a variation $A|\psi\rangle$ as
\[
  \mathrm{d} \Phi_{\mathrm{pure}}([\psi])(A|\psi\rangle) = 
  \bigl( \langle \psi | [H_1, A] | \psi \rangle, \dots, \langle \psi | [H_n, A] | \psi \rangle \bigr).
\]
The Jacobian has rank strictly less than $n$ if and only if the vectors $K_i |\psi\rangle$ are linearly dependent. That is, there exists a nonzero $\vec{\lambda} \in \mathbb{R}^n$ such that
\[
  \sum_{i=1}^n \lambda_i K_i |\psi\rangle = 0
  \quad \Leftrightarrow \quad
  \left( \sum_i \lambda_i H_i \right) |\psi\rangle = \lambda |\psi\rangle
\]
for some $\lambda \in \mathbb{R}$. Therefore:
\[
  \mathrm{rank}(\mathrm{d}\Phi_{\mathrm{pure}}([\psi])) < n
  \quad \Longleftrightarrow \quad
  [\psi] \in \Sigma.
\]

\paragraph{Full-rank locus is open and dense.}

Define the full-rank locus:
\[
  U := \mathbb{CP}^{d-1} \setminus \Sigma.
\]
Since $\Sigma$ has empty interior by assumption, $U$ is an open dense subset of $\mathbb{CP}^{d-1}$.  
On $U$, the differential $\mathrm{d}\Phi_{\mathrm{pure}}$ has full rank \(n\), so $\Phi_{\mathrm{pure}}$ is a submersion when restricted to $U$:
\[
  \Phi_{\mathrm{pure}}|_U \colon U \longrightarrow \mathbb{R}^n
\]
is smooth and has full-rank differential at every point.  
Hence its image
\[
  \Phi_{\mathrm{pure}}(U) \subset \mathbb{R}^n
\]
is an \emph{open subset} of $\mathbb{R}^n$ (by the submersion theorem).

\paragraph{Closure of the image.}

The map $\Phi_{\mathrm{pure}}$ is continuous and $U$ is dense in $\mathbb{CP}^{d-1}$, so:
\[
  \mathcal{M}
  = \Phi_{\mathrm{pure}}(\mathbb{CP}^{d-1})
  = \Phi_{\mathrm{pure}}(\overline{U})
  = \overline{\Phi_{\mathrm{pure}}(U)}.
\]
Therefore:
\[
  \mathcal{M} \text{ is the closure of an open set in } \mathbb{R}^n,
\]
and hence $\mathcal{M}$ is open in its affine span.

\paragraph{Convex hull and clopen argument.}

Since $\mathbb{CP}^{d-1}$ is compact and $\Phi_{\mathrm{pure}}$ is continuous, the image $\mathcal{M}$ is compact, and hence closed in $\mathbb{R}^n$. From Closure of the image, it is also open in its affine span.

But the convex hull $\overline{\mathcal{M}} = \mathrm{conv}(\mathcal{M})$ is convex (thus connected), and contains $\mathcal{M}$ as a nonempty subset that is both open and closed. Therefore:
\[
  \mathcal{M} = \overline{\mathcal{M}},
\]
as desired.
\end{proof}

\begin{remark}
\textbf{(Dimension bound)}  
Since $\Phi_{\mathrm{pure}}$ maps from a real manifold of dimension $2d - 2$ to $\mathbb{R}^n$, the rank of its Jacobian at any point is at most $2d - 2$. Therefore, full rank $n$ is only possible if
\[
  n \leq 2d - 2.
\]
This inequality is automatically satisfied on the open set $U = \mathbb{CP}^{d-1} \setminus \Sigma$ under the assumption that $\Sigma$ has empty interior.
\end{remark}

\begin{remark}
\textbf{(Genericity justification)}  
Thom’s Parametric Transversality Theorem implies that the condition $\Sigma$ has empty interior holds for a residual (i.e., generic) subset of $n$-tuples $(H_1,\dots,H_n)$ in the parameter space $\mathrm{Herm}(\mathcal{H})^n$. That is, the set of Hamiltonians for which $\mathcal{M} = \overline{\mathcal{M}}$ holds is topologically large (dense in the Baire category sense).
\end{remark}

\section{Equality of Ground-State QOS \(\mathcal{M}_0\) and Boundary \(\partial \mathcal{M}\)}

Let \(\mathcal{M} \subset \mathbb{R}^n\) denote the quantum observable space (QOS), defined as the set of all expectation values
\[
\vec{H} = \left( \langle H_1 \rangle, \ldots, \langle H_n \rangle \right)
\]
over pure quantum states. We assume \(\mathcal{M}\) is convex (not necessarily bounded above).

Let \(\mathcal{D} \subset S^{n-1}\) denote the set of directions \(\hat{\boldsymbol{\lambda}}\) such that the corresponding Hamiltonian
\[
H(\hat{\boldsymbol{\lambda}}) := \hat{\boldsymbol{\lambda}} \cdot \vec{H}
\]
is bounded from below. For all \(\hat{\boldsymbol{\lambda}} \in \mathcal{D}\), the ground state energy is finite and defined by
\[
E_0(\hat{\boldsymbol{\lambda}}) = \inf_{\vec{H} \in \mathcal{M}} \; \hat{\boldsymbol{\lambda}} \cdot \vec{H} \in \mathbb{R}.
\]
We assume the infimum is attained: there exists \(\vec{H}_0(\hat{\boldsymbol{\lambda}}) \in \mathcal{M}\) such that
\[
\hat{\boldsymbol{\lambda}} \cdot \vec{H}_0(\hat{\boldsymbol{\lambda}}) = E_0(\hat{\boldsymbol{\lambda}}).
\]
Define the ground state observable space as
\[
\mathcal{M}_0 := \left\{ \vec{H}_0(\hat{\boldsymbol{\lambda}}) \;\middle|\; \hat{\boldsymbol{\lambda}} \in \mathcal{D} \right\}.
\]
Then,
\[
\mathcal{M}_0 = \partial \mathcal{M},
\]
i.e., the set of ground state expectation values over all semibounded Hamiltonians coincides with the boundary of the observable space.

\begin{proof}
\emph{1. Every \(\vec{H}_0(\hat{\boldsymbol{\lambda}})\) lies on \(\partial \mathcal{M}\): } 

By definition of \(E_0(\hat{\boldsymbol{\lambda}})\),
\[
\hat{\boldsymbol{\lambda}} \cdot \vec{H} \ge E_0(\hat{\boldsymbol{\lambda}}) \quad \forall\, \vec{H} \in \mathcal{M},
\]
with equality at \(\vec{H}_0(\hat{\boldsymbol{\lambda}})\). Thus, the affine hyperplane
\[
H_{\hat{\boldsymbol{\lambda}}} := \left\{ \vec{y} \in \mathbb{R}^n \;\middle|\; \hat{\boldsymbol{\lambda}} \cdot \vec{y} = E_0(\hat{\boldsymbol{\lambda}}) \right\}
\]
supports \(\mathcal{M}\) at \(\vec{H}_0(\hat{\boldsymbol{\lambda}})\), so this point cannot lie in the interior. Hence \(\vec{H}_0(\hat{\boldsymbol{\lambda}}) \in \partial \mathcal{M}\).

\vspace{1em}
\noindent\emph{2. Every point on \(\partial \mathcal{M}\) arises as some \(\vec{H}_0(\hat{\boldsymbol{\lambda}})\):} 

Let \(\vec{p} \in \partial \mathcal{M}\). By the \emph{supporting hyperplane theorem}~\cite{boyd2004convex}, there exists a nonzero vector \(\boldsymbol{\lambda} \in \mathbb{R}^n\) and scalar \(c \in \mathbb{R}\) such that
\[
\boldsymbol{\lambda} \cdot \vec{p} = c, \quad \boldsymbol{\lambda} \cdot \vec{H} \ge c \quad \forall\, \vec{H} \in \mathcal{M}.
\]
Set \(\hat{\boldsymbol{\lambda}} = \boldsymbol{\lambda} / \|\boldsymbol{\lambda}\|\), so that \(\hat{\boldsymbol{\lambda}} \in \mathcal{D}\), and note that
\[
\hat{\boldsymbol{\lambda}} \cdot \vec{p} = \frac{c}{\|\boldsymbol{\lambda}\|} \le \inf_{\vec{H} \in \mathcal{M}} \hat{\boldsymbol{\lambda}} \cdot \vec{H} = E_0(\hat{\boldsymbol{\lambda}}),
\]
but since \(\vec{p} \in \mathcal{M}\), we also have \(E_0(\hat{\boldsymbol{\lambda}}) \le \hat{\boldsymbol{\lambda}} \cdot \vec{p}\). Therefore,
\[
\hat{\boldsymbol{\lambda}} \cdot \vec{p} = E_0(\hat{\boldsymbol{\lambda}}),
\]
and \(\vec{p}\) achieves the infimum. By definition, \(\vec{p} = \vec{H}_0(\hat{\boldsymbol{\lambda}})\), and so \(\vec{p} \in \mathcal{M}_0\).
\end{proof}

\section{Gaussian Curvature for QOS of Kitaev Honeycomb Model}
In this section, we derive an analytical expression for the Gaussian curvature of the quantum observable space (QOS) associated with the Kitaev honeycomb model. We show that the curvature vanishes precisely at the quantum phase boundary.

\subsection{Gaussian Curvature for 1-Homogeneous Functions}
Let \( E_0 : \mathbb{R}^{n} \setminus \{0\} \to \mathbb{R} \) be a smooth function that is homogeneous of degree 1. Define the Hessian matrix
\[
\H := \nabla^2 E(\vec{J}).
\]
Let \( S \subset \mathbb{R}^n \setminus \{0\} \) be an \((n-1)\)-dimensional submanifold such that \( \vec{J} \notin T_{\vec{J}}S \) for all \( \vec{J} \in S \). Define the image surface
\[
\mathcal{S} := \bigl\{ \vec{H}(\vec{J}) = \nabla E(\vec{J}) \mid \vec{J} \in S \bigr\} \subset \mathbb{R}^n.
\]

By Euler's theorem for homogeneous functions,
\[
E_0(\vec{J}) = \vec{J} \cdot \nabla E_0(\vec{J})
\quad \Rightarrow \quad
\vec{J} \cdot \H = \vec{0},
\]
so \( \H \) has rank at most \( n-1 \), with null space spanned by \( \vec{J} \). Since \( \vec{J} \notin T_{\vec{J}}S \), the orthogonal projection
\[
P := I - \hat{J} \hat{J}^\mathsf{T},
\quad \text{with} \quad
\hat{J} := \frac{\vec{J}}{\|\vec{J}\|},
\]
projects onto the \((n-1)\)-dimensional subspace orthogonal to \( \vec{J} \), and \( P\H P \) restricts \( \H \) to its non-null block. Thus, the Gaussian curvature of the surface \( \mathcal{S} \) is given by
\[
K_S(\vec{J}) = \frac{1}{\det(P \H P)} = \frac{1}{\mathrm{pdet}(\H)},
\]
where \( \mathrm{pdet}(\H) \) denotes the pseudo-determinant (product of nonzero eigenvalues). For \( n = 3 \), where \( \H \) generically has rank 2, we have
\begin{equation}
\mathrm{pdet}(\H) = \frac{1}{2} \left[ \left( \tr \H \right)^2 - \tr(\H^2) \right].
\label{eq:rank2}
\end{equation}

\subsection{Ground State Energy of the Kitaev Model}

The Hamiltonian of the Kitaev honeycomb model is
\begin{equation}
H = -J_x \sum_{\langle ij \rangle_x} \sigma_i^x \sigma_j^x 
    - J_y \sum_{\langle ij \rangle_y} \sigma_i^y \sigma_j^y 
    - J_z \sum_{\langle ij \rangle_z} \sigma_i^z \sigma_j^z,
    \notag
\end{equation}
where \( J_x, J_y, J_z \) denote anisotropic couplings along the three bond directions of the honeycomb lattice~\cite{kitaev2006anyons,feng2007topological,chen2008exact}.

The ground-state energy per unit cell is given by
\begin{equation}
E_0 = -\frac{1}{\mathrm{Vol}(\text{BZ})} \int_{\text{BZ}} \frac{d^2 \vec{k}}{(2\pi)^2} \, |f(\vec{k})|,
\label{eq:ground}
\end{equation}
where:
\begin{itemize}
  \item The integration is over the first Brillouin zone (BZ).
  \item The factor of 2 from the Majorana fermion representation has been absorbed.
  \item \( f(\vec{k}) := \vec{J} \cdot \vec{\Phi}(\vec{k}) \), with
  \[
  \vec{\Phi}(\vec{k}) := \left( e^{i \vec{k} \cdot \vec{a}_x}, \, e^{i \vec{k} \cdot \vec{a}_y}, \, 1 \right),
  \]
  where \( \vec{a}_x, \vec{a}_y \) are Bravais lattice vectors.
\end{itemize}

\subsection{Hessian of the Ground State Energy}

From Eq.~\eqref{eq:ground}, the Hessian matrix is given by
\[
\H = -\frac{1}{\mathrm{Vol}(\text{BZ})} \int_{\text{BZ}} \frac{d^2 \vec{k}}{(2\pi)^2} \, h(\vec{k}),
\]
with \( h(\vec{k}) := \nabla^2_{\vec{J}} |f(\vec{k})| \). Using \( \nabla_{\vec{J}} f = \vec{\Phi} \), we compute:
\[
\nabla_{\vec{J}} |f| = \frac{1}{|f|} \Re(f^* \vec{\Phi}),
\]
and
\[
\nabla_{\vec{J}}^2 |f| = \frac{1}{|f|^3} \left( \Re(f^* \vec{\Phi} \otimes f \vec{\Phi}^*) - \Re(f^* \vec{\Phi}) \otimes \Re(f^* \vec{\Phi}) \right).
\]
Using the identity
\[
\Re(a) \otimes \Re(a) + \Im(a) \otimes \Im(a) = \Re(a \otimes a^*),
\]
we obtain the compact expression:
\[
h(\vec{k}) = \frac{1}{|f|^3} \, \Im(f^* \vec{\Phi}) \otimes \Im(f^* \vec{\Phi}) = \vec{u}(\vec{k}) \otimes \vec{u}(\vec{k}),
\]
where
\[
\vec{u}(\vec{k}) := \frac{\Im(f^* \vec{\Phi})}{|f|^{3/2}}.
\]
Thus,
\[
\H = -\frac{1}{\mathrm{Vol}(\text{BZ})} \int_{\text{BZ}} \vec{u}(\vec{k}) \vec{u}(\vec{k})^\mathsf{T} \, d^2 \vec{k}.
\]
This is a sum of rank-1 matrices, and the resulting \( \H \) generically has rank 2.

Using Eq.~\eqref{eq:rank2}, the pseudo-determinant is
\begin{equation}
\mathrm{pdet}(\H) = \frac{1}{2\, \mathrm{Vol}(\text{BZ})^2} \iint_{\text{BZ}} \left( \vec{u}(\vec{k}) \wedge \vec{u}(\vec{k}') \right)^2 \, d^2 \vec{k} \, d^2 \vec{k}'.
\end{equation}

\subsection{Divergence at the Phase Boundary}

We now analyze the divergence of \( \H \) at the phase boundary of the Kitaev model. The key distinction lies in the low-energy dispersion near the Dirac points.

\paragraph{Inside the Gapless Phase.} The function \( f(\vec{k}) \) has two isolated zeros in the BZ (Dirac points \( \vec{k}_D \)). Expanding near a Dirac point:
\[
f(\vec{k}_D + \vec{q}) \approx \nabla f|_{\vec{k}_D} \cdot \vec{q} + \cdots,
\quad \Rightarrow \quad
\epsilon(\vec{k}) \approx |\vec{q}|,
\]
so
\[
\partial_J^2 \epsilon(\vec{k}) \sim \frac{1}{|\vec{q}|}, \quad
\Rightarrow \quad \H \sim \int \frac{d^2 q}{|\vec{q}|} \sim \int dq,
\]
which is finite due to compactness of the BZ.

\paragraph{At the Phase Boundary.} The Dirac points merge at a single point \( \vec{k}_0 \), where
\[
f(\vec{k}_0) = 0, \quad \nabla f|_{\vec{k}_0} = 0.
\]
Near \( \vec{k}_0 \), we expand:
\[
f(\vec{k}_0 + \vec{q}) \approx A(q_x^2 - q_y^2) + B q_x q_y + \cdots \quad \Rightarrow \quad \epsilon(\vec{k}) \sim |\vec{q}|^2.
\]
If \( A = A(\vec{J}) \to 0 \) as \( \vec{J} \to \vec{J}_c \), then
\[
\epsilon(\vec{k}) \sim A(\vec{J}) |\vec{k}|^2,
\quad \Rightarrow \quad
\partial_J^2 \epsilon(\vec{k}) \sim \frac{1}{|\vec{k}|^2},
\]
so the Hessian diverges logarithmically:
\[
\H \sim \int \frac{d^2 k}{|\vec{k}|^2} \sim \log\left( \frac{1}{k_{\text{IR}}} \right),
\]
due to the infrared divergence as \( k \to 0 \). Hence, the Gaussian curvature \( K_S(\vec{J}) \sim 1/\mathrm{pdet}(\H) \) vanishes at the phase transition.

\section{Quantum Potts Model}

As a representative example of a Type-I transition in our framework, we consider the quantum \( q \)-state Potts model. It is well known that this model exhibits a first-order quantum phase transition for \( q \geq 5 \) in \(1+1\) dimensions and for \( q \geq 3 \) in \(2+1\) dimensions~\cite{solyom1981renormalization}. Here, we focus on the \(2+1\)D three-state Potts model, using numerical data extracted directly from Fig.~5 of Ref.~\cite{ding2017monte}.

\begin{figure}[ht]
\centering
\includegraphics[width=1\linewidth]{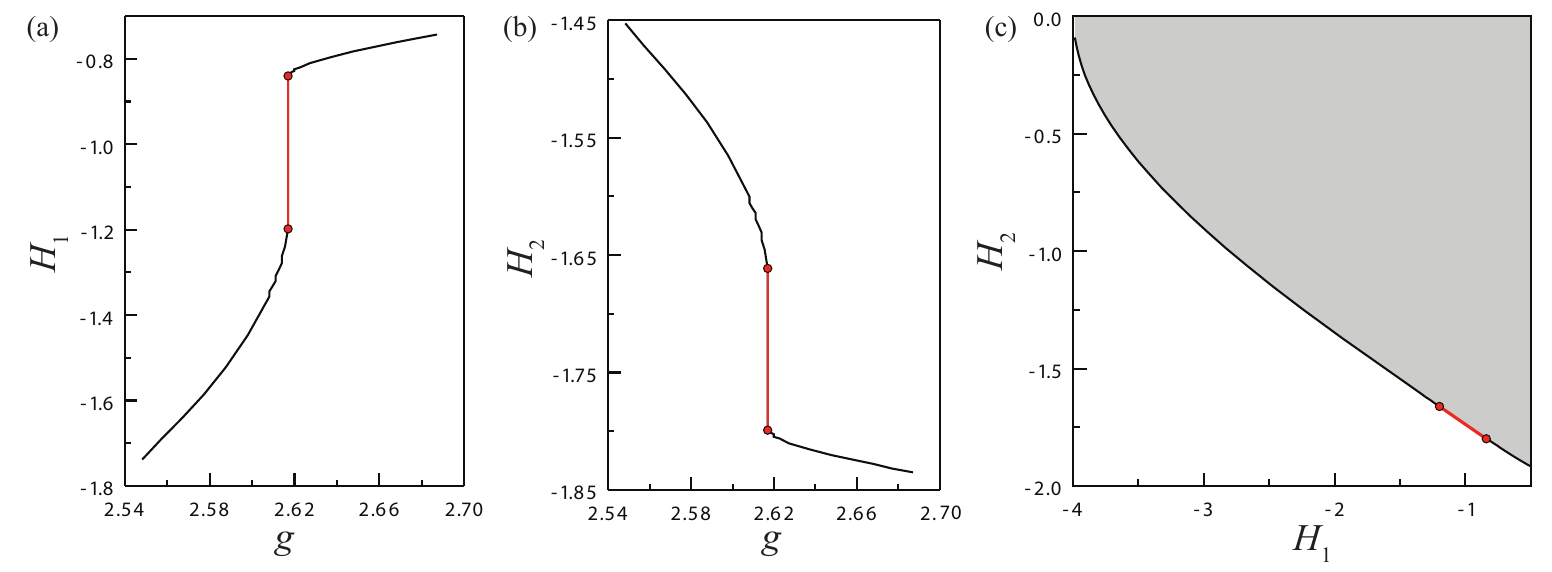}
\caption{
Analysis of the \(2+1\)D three-state Potts model: (a) Expectation value \( \langle H_1 \rangle \) vs.\ coupling \( g \); (b) Expectation value \( \langle H_2 \rangle \) vs.\ \( g \); both exhibit sharp discontinuities at the first-order transition near \( g \approx 2.62 \). (c) The corresponding QOS, where the red circles mark the transition points and the red line indicates the theoretical coexistence region, consistent with a Type-I singularity.
}
\label{fig:potts}
\end{figure}

As shown in Fig.~\ref{fig:potts}(a)--(b), a first-order transition occurs near \( g \approx 2.62 \), evidenced by discontinuities in both \( \langle H_1 \rangle \) and \( \langle H_2 \rangle \). These points, marked by red circles, correspond to distinct ground states on either side of the transition. The corresponding QOS is plotted in Fig.~\ref{fig:potts}(c), where the gap between the red circles indicates a non-analyticity of the image under the expectation map, i.e., a Type-I singularity.

Although the intermediate states are not directly sampled in the Monte Carlo simulation, the convexity of the QOS implies the existence of a quantum coexistence region connecting the two degenerate ground states. This region corresponds to convex combinations (or superpositions, at zero temperature) of the extremal states and is illustrated by the red line in Fig.~\ref{fig:potts}(c).

\end{document}